\documentclass{article}
\PassOptionsToPackage{numbers,sort&compress}{natbib} 
\usepackage[final,sglblindworkshop]{neurips_2025}
\workshoptitle{Efficient Reasoning}

\usepackage[utf8]{inputenc}
\usepackage[T1]{fontenc}
\usepackage{hyperref}
\usepackage{url}
\usepackage{booktabs}       
\usepackage{amsmath,amssymb,amsfonts}
\usepackage{microtype}
\usepackage{graphicx}
\usepackage{xcolor}
\usepackage{enumitem}
\usepackage[most]{tcolorbox} 
\tcbset{
  enhanced,
  breakable,
  boxrule=0.4pt,
  colback=gray!6,
  colframe=gray!40,
  arc=2mm,
  left=6pt,
  right=6pt,
  top=6pt,
  bottom=6pt,
  before skip=6pt,
  after skip=6pt
}

\newtcolorbox{promptbox}[1][]{
  enhanced,
  breakable,
  colback=gray!6,
  colframe=gray!40,
  coltitle=black,
  colbacktitle=gray!10,
  fonttitle=\bfseries,
  boxrule=0.4pt,
  arc=2mm,
  left=8pt, right=8pt, top=8pt, bottom=8pt,
  before skip=6pt, after skip=6pt,
  #1
}

\setlist[itemize]{leftmargin=*, itemsep=1pt, topsep=2pt}
\setlist[enumerate]{leftmargin=*, itemsep=1pt, topsep=2pt}

\usepackage{multirow}
\usepackage[table]{xcolor}

\title{Reuse, Don’t Recompute: Efficient Large Reasoning Model Inference via Memory Orchestration}

\author{
  Daivik Patel\thanks{Equal contribution. Correspondence to daivik.d.patel@rutgers.edu and shrenik.d.patel@rutgers.edu.} \\
  \ \ \ Rutgers University \\
  \And
  Shrenik Patel\footnotemark[1] \\
  Rutgers University \\
}

\begin{document}

\maketitle

\begin{abstract}
Large reasoning models (LRMs) achieve strong accuracy through test-time scaling (TTS), generating longer chains of thought or sampling multiple solutions, but at steep costs in tokens and latency. We argue that memory is a core ingredient for efficient reasoning: when evidence already exists, models should “think less” by reusing structured memory instead of recomputing derivations. We present ENGRAM-R, an inference-time memory layer that integrates typed retrieval with compact fact card representations and explicit citation control. On the LoCoMo benchmark, ENGRAM-R reduces input tokens by 85\% and reasoning tokens by 75\% versus full context while maintaining high accuracy. On a multi-hop slice of the LongMemEval benchmark, it achieves similar efficiency with substantial accuracy gains. These results show that memory is not only critical for long-horizon correctness, but also a practical lever for efficient reasoning under tight compute, memory, and latency budgets.
\end{abstract}

\section{Introduction}
Large reasoning models (LRMs) have made “thinking longer” a default recipe for better answers: expand the chain-of-thought (CoT) \citep{Wei2022CoT}, sample more drafts \citep{Wang2023SelfConsistency}, and hope consensus emerges. Although this principle yields in real gains, it doesn’t come without costs. Lengthy test-time reasoning inflates latency, burns tokens on evidence the model already “knows,” and turns deployment into an exercise in budget triage rather than engineering. If LRMs are to be used in constrained settings, we need a way \emph{to spend fewer tokens without compromising accuracy}.

We argue that the right lever is not \emph{more} thought, but \emph{less redundant} thought. Focusing on test-time compute, we set the goal of reducing the tokens and wall-time an LRM spends during inference. Our premise is simple: many tasks (especially those with recurring entities, facts, and routines) don’t require recomputing long chains each time. Instead, they benefit from \textbf{typed, reusable evidence} that can be selectively retrieved and composed at inference. We operationalize this with a \textbf{memory-orchestrated inference} layer that sits strictly outside the model weights. It organizes interaction traces and knowledge as \textbf{episodic, semantic, and procedural} records; performs \textbf{dense, type-aware retrieval} into a small fixed budget; and ultimately helps \textbf{constrain a model's reasoning}.

Our contributions are the following. (1) We present ENGRAM-R, a \textbf{memory-orchestrated inference} recipe for LRMs that empirically reduces test-time cost without touching model weights. (2) We deliver evidence on \textbf{two long-horizon benchmarks} that typed retrieval can replace large fractions of CoT while preserving fidelity, along with a clear diagram and reasoning trace to make the process auditable. The message is clear and actionable: \emph{test-time scaling is not the only path to reliability}. For many real deployments, \textbf{reuse beats recompute}. Typed memory and selective retrieval let LRMs reach correctness sooner, with fewer tokens, lower latency, and tighter budgets.

\section{Related Work}
Research on long-term memory for language models spans non-parametric retrieval, structured graphs, and system-level abstractions. Early work coupled frozen models with dense or lexical retrieval \citep{Khandelwal2020,Lewis2020,Guu2020}, improving factual recall but often relying on heuristic chunking and calibration. Retrieval-pretrained and nearest-neighbor approaches scale this line further \citep{Borgeaud2022}, trading simplicity for freshness and editability. Structured methods instead frame memory as a graph, capturing entities and relations to support multi-hop reasoning \citep{Anokhin2024,Li2024GraphReader}, while system-level proposals treat memory as a schedulable resource with paging and lifecycle management \citep{Packer2023,Li2025MemOS}. These designs extend capacity but typically return verbose snippets, leaving models to repeatedly re-narrate evidence.

Parallel to advances in memory, work on LRMs has begun to emphasize efficiency. While early advances in LRMs emphasized thinking longer through extended chains or ensemble-style decoding, more recent work has shifted toward making inference more efficient. Methods such as \emph{Skeleton-of-Thought} \citep{Ning2023SoT} and \emph{Chain-of-Draft} \citep{Xu2025CoD} restructure reasoning to reduce redundancy and accelerate generation, showing that careful control over inference can yield accuracy at lower cost. Our approach is complementary: rather than restructuring the chain itself, we compress the evidence surface, reducing the need for redundant reasoning by constraining what models are allowed to see and cite.

By re-rendering retrieved content into compact, traceable \emph{Fact Cards}, ENGRAM-R enables LRMs to handle long-horizon reasoning with \textbf{sharply reduced token and latency budgets} while maintaining competitive accuracy—a practical path to efficient reasoning at scale.

\section{The ENGRAM-R Architecture}
\label{sec:method}
\begin{figure}[h!]
\centering
\includegraphics[width=.98\linewidth]{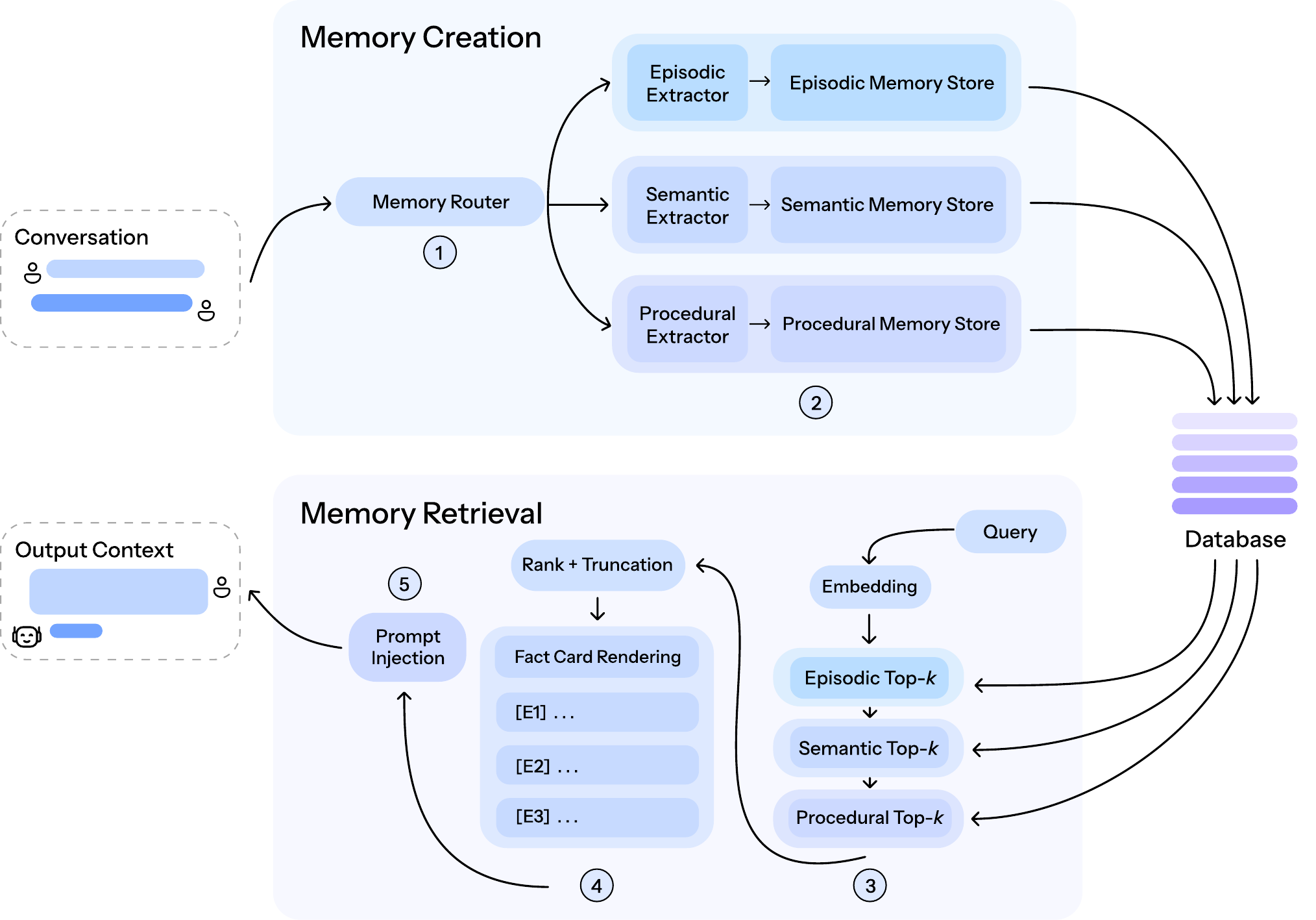}
\caption{\textbf{System overview of ENGRAM-R.} Dialogue turns are routed into typed stores (episodic, semantic, procedural), embedded, and retrieved at query time. Retrieved items are re-rendered into compact \emph{Fact Cards}, which provide atomic, anchored claims for direct citation. This design bounds input size, shortens reasoning, and enables LRMs to achieve efficient long-horizon inference. Numbers (1)–(5) mark the main components and are referenced below.}
\label{fig:engram-overview}
\end{figure}

We extend the ENGRAM memory architecture \citep{patel2025engram} into a full \textbf{efficiency layer for LRMs}. Analogous to ENGRAM, our system organizes dialogue into typed memory stores \citep{Tulving1972,Cohen1980} and performs dense retrieval at query time, but we add two critical extensions: (i) retrieved items are re-rendered into compact \emph{Fact Cards}, and (ii) models are instructed to \emph{cite cards directly} in their reasoning trace. These additions turn memory from a verbose context dump into a bounded, auditable evidence substrate. The pipeline consists of five main stages, marked (1)–(5) in Figure~\ref{fig:engram-overview} and are referenced throughout this section. Here, we briefly recap the essentials and then introduce our extensions. 

At a high level, ENGRAM routes each dialogue turn into one or more typed stores (episodic, semantic, procedural) (1), normalizes it into a lightweight record paired with an embedding, and persists it in a relational memory (2). At query time, the system retrieves top-$k$ candidates from each store using cosine similarity, merges and deduplicates them, and truncates to a fixed budget (typically $K{=}25$) (3). A full specification of the original ENGRAM architecture we extend is in Appendix~\ref{app:engram-base}.

\subsection{Problem Setup}
We study the problem of \emph{efficient long-horizon reasoning} in LRMs.  
In this setting, a dialogue unfolds turn by turn, with each entry consisting of a speaker, their dialogue, and a timestamp. At query time, the model must answer a question that may depend on information from any point in the conversation, even far back in history.  

A naive baseline is to pass the \emph{entire conversation} as input. This guarantees recall but scales poorly: tokens grow linearly with turns, often exceeding context limits. Test-time scaling (TTS) strategies such as extended chains \citep{Wei2022CoT} or self-consistency \citep{Wang2023SelfConsistency} amplify cost further, inflating both latency and token budget. To avoid this, ENGRAM uses retrieval:  
\begin{equation}
\tilde{R}(q) \;=\; \operatorname{TopK}\{\, m \in \mathcal{M} \mid \mathrm{score}(q,m)\,\},
\end{equation}
where $\mathcal{M}$ is the typed memory state, $\mathrm{score}$ is a similarity function, and $\tilde{R}(q)$ is a compact set of candidates that ensures coverage without replaying the full dialogue.  

The limitation is that even $\tilde{R}(q)$ contains multi-sentence, verbose snippets. Passing these to an LRM forces it to re-narrate evidence before using it, wasting tokens and elongating reasoning chains. Our contribution is to transform $\tilde{R}(q)$ into compact, transparent \emph{Fact Cards}, paired with a controlled citation mechanism, enabling LRMs to consume retrieval efficiently and within bounded reasoning cost.  

\subsection{Extensions to ENGRAM}

Our contribution extends this pipeline at two key stages:  

\paragraph{1. Aggregation $\rightarrow$ Fact Card Rendering.}  
Rather than injecting raw, often verbose record text, we re-render each retrieved memory $m \in \tilde{R}(q)$ into a compact \textbf{Fact Card} (4). A Fact Card captures only the essentials of a record  
\[
\phi(m) \;=\; \big[\,\text{id}(m),\; \text{claim}(m),\; \text{anchor}(m)\,\big]
\]  
where $\text{id}(m)$ is a stable identifier (e.g., [E1], [E2]), $\text{claim}(m)$ is a minimal canonical statement distilled from the record, and $\text{anchor}(m)$ is a timestamp or provenance marker that grounds the claim. Formally, the set of retrieved records $\tilde{R}(q)$ is transformed into a new evidence state  

\[
\mathcal{F}(q) \;=\; \{\phi(m) \mid m \in \tilde{R}(q)\}.
\]  

Intuitively, $\mathcal{F}(q)$ is a set of atomic, non-redundant claims tied to explicit provenance. This step compresses multi-sentence snippets into concise, verifiable evidence units that the model can cite directly, while preserving the information required for faithful reasoning.

\paragraph{2. Prompt Construction $\rightarrow$ Controlled Citation.}  
In the final stage, the answering model is conditioned not on raw text but on the compact set of Fact Cards $\mathcal{F}(q)$ (5). To ensure accountability, we add a \emph{controlled citation mechanism}: the model must reference evidence by explicit identifiers rather than paraphrasing. Formally, the prompt is constructed as  
\[
P(q) \;=\; \mathrm{Template}\!\big(q, \mathcal{F}(q)\big)
\]  
where the template specifies that answers must cite card IDs (e.g., [E1], [E2], …) when justifying claims. At inference, the generated answer $\hat{a}$ is required to satisfy  

\[
\hat{a} \;\Rightarrow\; \big\{\, [E_i] \;\big|\; \phi(m_i) \in \mathcal{F}(q) \,\big\},
\]  

meaning that any citation in $\hat{a}$ must correspond to a valid Fact Card in the retrieved set. This constraint bounds reasoning length by preventing the model from re-describing evidence, while also making outputs \emph{auditable}: every justification can be traced back to a specific, compact Fact Card.

\paragraph{Overview.}  
Together, Fact Cards and controlled citation reframe ENGRAM from a generic memory module into an \emph{efficiency layer for LRMs}. By compressing evidence into $\phi(m)$ and constraining its use through explicit citation, we reduce both input and reasoning cost while maintaining fidelity. Importantly, this provides a transparent efficiency layer: answers are short, accountable, and directly linked to their supporting records. For a step-by-step walkthrough of this process, see Appendix~\ref{app:walkthrough}. 

\section{Evaluation}
We evaluate ENGRAM-R on two complementary long-horizon conversational benchmarks. LoCoMo compresses realistic two-speaker dialogues into long, multi-session conversations spanning diverse reasoning categories; LongMemEval$_\text{S}$ embeds QA in extended user–assistant histories that stress multi-session and temporal reasoning. Our evaluation protocol reports judge-based answer quality alongside input/reasoning tokens and p50/p95 latency. Numerical results are detailed in the next section.

\subsection{Benchmarks}
\textbf{LoCoMo.} LoCoMo consists of multi-session dialogues built through a human–machine pipeline grounded in personas and event graphs, then refined by human annotators for long-range consistency \citep{Maharana2024}. It contains 10 dialogues, each averaging 600 turns and $\approx$16K tokens across up to 32 sessions. The QA split covers five categories: single-hop, multi-hop, temporal, open-domain, and adversarial. In line with prior work, we exclude adversarial cases when reporting QA metrics and present category-level results for the remaining four. \\ \\
\textbf{LongMemEval$_\text{S}$.} LongMemEval is designed to test interactive memory in user–assistant dialogues, covering five abilities: information extraction, multi-session reasoning, temporal reasoning, knowledge updates, and abstention (declining to answer when evidence is insufficient) \citep{Wu2024LongMemEval}. It contains 500 curated questions embedded in chat histories of configurable length. For our study, we focus on LongMemEval$_\text{S}$ ($\approx$115K tokens per problem) and report QA metrics on the multi-session reasoning and temporal reasoning categories. These settings are particularly challenging and better highlight whether memory orchestration can deliver efficiency gains without sacrificing accuracy.

\subsection{Backbone and baselines}
All experiments are conducted with \texttt{gpt-oss-20b} \citep{gptoss2025}, selected as a strong open-source LRM with competitive chain-of-thought capabilities. The model weights remain frozen; our interventions act only at inference time. We evaluate two inference settings: 

\begin{enumerate}[label=\textbf{\arabic*.}]
    \item \textbf{Full-Context baseline.} The model is provided with the complete dialogue history for each query and allows unconstrained reasoning, representing the standard LRM-only deployment approach.
    
    \item \textbf{ENGRAM-R.} This setting augments inference with a compact memory orchestration layer: interaction traces and knowledge are organized into typed records; for a given query, we perform dense retrieval and truncate to a fixed evidence budget \( K \) before presenting this condensed context to the model. The memory layer runs with \texttt{gpt-4o-mini} \citep{OpenAI2023}, a non-reasoning, intentionally fast and lightweight model, as its backbone, conserving reasoning for the question-answering stage.
\end{enumerate}

\subsection{Metrics}

To evaluate the performance of our approach, we consider two primary dimensions that capture both the quality of the output and the computational efficiency of the system:

\begin{enumerate}[label=\textbf{\arabic*.}]
    \item \textbf{Quality.} We use an \textbf{LLM-as-Judge} protocol \citep{Zheng2023LLMJudge} as the main metric of semantic correctness. This involves using an independent LLM that, given the question, gold answer, and prediction, renders a binary semantic-correctness decision based on factual fidelity, completeness, and contextual appropriateness of the response. This ensures robustness to surface-level variation in responses.
    
    \item \textbf{Efficiency.} We decompose efficiency into:
    \begin{itemize}
        \item \textbf{Input tokens:} the number of tokens supplied to the model (prompt + retrieved evidence), reflecting context compression.
        \item \textbf{Reasoning tokens:} the number of tokens generated as chain-of-thought, reflecting the cost of inference-time reasoning.
        \item \textbf{Latency:} reported as end-to-end wall-clock time per query with p50 and p95 to capture both typical and tail behavior.
    \end{itemize}
\end{enumerate}

These metrics provide a comprehensive assessment of both the model's reasoning capabilities and its operational efficiency. By examining quality alongside efficiency, we gain a deeper understanding of how well ENGRAM-R balances performance with computational cost, offering valuable insights for real-world deployment scenarios.

\section{Results}
\label{results}
In this section, we present a comprehensive evaluation of ENGRAM-R across multiple datasets and reasoning tasks. The results highlight how the proposed approach consistently improves both efficiency and accuracy, particularly in scenarios requiring long-horizon reasoning, without sacrificing performance.

\begin{table*}[h!]
\centering
\small
\setlength{\tabcolsep}{6pt}
\caption{\textbf{LoCoMo results}: input tokens, reasoning tokens, and judge accuracy across QA categories. ENGRAM-R yields large efficiency gains while improving multi-hop and temporal accuracy.\\}
\label{tab:locomo}
\begin{tabular}{l l r r r}
\toprule
\textbf{Category} & \textbf{Setting} & \textbf{Input tokens} & \textbf{Reasoning tokens} & \textbf{LLM judge (\%)} \\
\midrule
\multirow{2}{*}{Single-hop}
  & Full-Context & 15{,}614{,}211 & 698{,}845 & 84.7 \\
  & \cellcolor{gray!9}ENGRAM-R
    & \cellcolor{gray!9}1{,}802{,}531
    & \cellcolor{gray!9}199{,}718
    & \cellcolor{gray!9}79.1 \\
\midrule
\multirow{2}{*}{Multi-hop}
  & Full-Context &  5{,}187{,}624 & 300{,}631 & 72.0 \\
  & \cellcolor{gray!9}ENGRAM-R
    & \cellcolor{gray!9}602{,}634
    & \cellcolor{gray!9}81{,}035
    & \cellcolor{gray!9}74.5 \\
\midrule
\multirow{2}{*}{Temporal}
  & Full-Context &  5{,}786{,}564 & 271{,}193 & 67.3 \\
  & \cellcolor{gray!9}ENGRAM-R
    & \cellcolor{gray!9}686{,}947
    & \cellcolor{gray!9}74{,}337
    & \cellcolor{gray!9}69.2 \\
\midrule
\multirow{2}{*}{Open-domain}
  & Full-Context &  1{,}783{,}304 &  65{,}319 & 64.6 \\
  & \cellcolor{gray!9}ENGRAM-R
    & \cellcolor{gray!9}201{,}366
    & \cellcolor{gray!9}23{,}334
    & \cellcolor{gray!9}57.2 \\
\midrule
\multirow{2}{*}{Overall}
  & Full-Context & 28{,}371{,}703 & 1{,}335{,}988 & 77.5 \\
  & \cellcolor{gray!9}\textbf{ENGRAM-R}
    & \cellcolor{gray!9}\textbf{3{,}293{,}478}
    & \cellcolor{gray!9}\textbf{378{,}424}
    & \cellcolor{gray!9}\textbf{75.6} \\
\bottomrule
\end{tabular}
\end{table*}

\subsection{LoCoMo: Category-level Analysis}
Table~\ref{tab:locomo} reports LoCoMo results by QA category. The data conveys how ENGRAM-R consistently \textbf{reduces inputs by $\approx$89\%} across categories and \textbf{shrinks reasoning token usage by $\approx$72\%}. Crucially, accuracy is \emph{maintained or improved} precisely where long-range composition matters most: \textbf{multi-hop (+2.5\%)} and \textbf{temporal (+1.9\%)}. These gains support the hypothesis that typed memory plus compact, citable evidence reduces the model's need to re-derive multi-step chains already present in the dialogue history. It is also evident that two categories show accuracy headroom. Single-hop (short, localized queries) experiences a slight decrease, which aligns with the idea that heavy truncation may not be necessary for questions focused on immediate context. The largest difference appears in open-domain settings, where having access to broader background knowledge would potentially be useful.

\begin{table*}[h!]
\centering
\small
\setlength{\tabcolsep}{6pt}
\caption{\textbf{LongMemEval$_\text{S}$ results}: compact typed evidence improves efficiency and accuracy on very long histories.\\}
\label{tab:lme}
\begin{tabular}{l l r r r}
\toprule
\textbf{Category} & \textbf{Setting} & \textbf{Input tokens} & \textbf{Reasoning tokens} & \textbf{LLM judge (\%)} \\
\midrule
\multirow{2}{*}{Multi-session}
  & Full-Context & 13{,}741{,}996 &  94{,}915 & 36.8 \\
  & \cellcolor{gray!9}ENGRAM-R
    & \cellcolor{gray!9}631{,}083
    & \cellcolor{gray!9}20{,}920
    & \cellcolor{gray!9}66.9 \\
\midrule
\multirow{2}{*}{Temporal}
  & Full-Context & 13{,}741{,}916 & 154{,}180 & 39.1 \\
  & \cellcolor{gray!9}ENGRAM-R
    & \cellcolor{gray!9}602{,}173
    & \cellcolor{gray!9}33{,}831
    & \cellcolor{gray!9}52.6 \\
\midrule
\multirow{2}{*}{Overall}
  & Full-Context & 27{,}483{,}912 & 245{,}125 & 38.0 \\
  & \cellcolor{gray!9}\textbf{ENGRAM-R}
    & \cellcolor{gray!9}\textbf{1{,}233{,}256}
    & \cellcolor{gray!9}\textbf{54{,}301}
    & \cellcolor{gray!9}\textbf{59.8} \\
\bottomrule
\end{tabular}
\end{table*}

\subsection{LongMemEval$_\text{S}$: Performance Analysis}
Table~\ref{tab:lme} summarizes performance on the LongMemEval$_\text{S}$ slice that includes the multi-session and temporal reasoning categories. Full-context accuracy is low despite unbounded inputs, suggesting that extremely long histories hinder effective use of context in LRMs. ENGRAM-R reverses this: by condensing and typing prior information into compact, auditable evidence, the model avoids being lost in a sea of context (100k+ tokens) and instead is able to cite what matters. The result is a \textbf{+21.8\% overall accuracy} improvement alongside \textbf{$\approx$96\% fewer input tokens} and \textbf{$\approx$78\% fewer reasoning tokens}. As mentioned, the gains are incredibly evident in both the \textbf{multi-session (+30.1)} and \textbf{temporal reasoning (+13.5)} categories, where durable memory and temporal anchors are essential for efficient reasoning. 

\begin{table*}[h!]
\centering
\small
\setlength{\tabcolsep}{8pt}
\caption{\textbf{Search and total inference latency (seconds).} We report median (p50) and tail (p95) latencies across datasets. 
Dashes indicate that search latency is not applicable in the full-context baseline.\\}
\label{tab:latency}
\begin{tabular}{l l cc cc}
\toprule
\textbf{Dataset} & \textbf{Setting} & \multicolumn{2}{c}{\textbf{Search}} & \multicolumn{2}{c}{\textbf{Total}} \\
\cmidrule(lr){3-4}\cmidrule(lr){5-6}
 &  & p50 & p95 & p50 & p95 \\
\midrule
\multirow{2}{*}{LoCoMo}
  & Full-Context & --   & --   & 7.89 & 17.16 \\
  & \cellcolor{gray!9}\textbf{ENGRAM\text{-}R}
    & \cellcolor{gray!9}\textbf{1.04} & \cellcolor{gray!9}\textbf{2.67}
    & \cellcolor{gray!9}\textbf{2.56} & \cellcolor{gray!9}\textbf{7.43} \\
\midrule
\multirow{2}{*}{LongMemEval$_\text{S}$}
  & Full-Context & --   & --   & 9.62 & 21.47 \\
  & \cellcolor{gray!9}\textbf{ENGRAM\text{-}R}
    & \cellcolor{gray!9}\textbf{0.72} & \cellcolor{gray!9}\textbf{1.18}
    & \cellcolor{gray!9}\textbf{1.88} & \cellcolor{gray!9}\textbf{5.54} \\
\bottomrule
\end{tabular}
\end{table*}

\subsection{Latency and Tail Behavior}

Latency trends in both methods follow token reductions (Table~\ref{tab:latency}). Relative to full-context, ENGRAM-R \textbf{reduces median total latency} during QA time by \textbf{$\approx$68\%} on LoCoMo and \textbf{$\approx$81\%} on LongMemEval$_\text{S}$. Tail latency (p95) sees improvements on a similar magnitude. The memory search stage itself is fast, with p50 times of $\le$ 1.04\,s, indicating that the retrieval process does not introduce much overhead. The primary efficiency gains, however, come from the reduction in both the input context size and the reasoning tokens generated during the answering process. By limiting the model’s input and focusing on the most relevant information, ENGRAM-R effectively shrinks the reasoning footprint, leading to faster response times.

\subsection{Summary and Implications}
ENGRAM-R consistently delivers \textbf{order-of-magnitude input compression and large reductions in reasoning tokens}. Across datasets and metrics, the findings support the effectiveness of compact, typed memory as a means to \textbf{reduce test-time compute} in LRMs without compromising fidelity on tasks requiring long-horizon reasoning. ENGRAM-R maintains strong accuracy on LoCoMo while materially reducing both the context ingested and the reasoning effort required. On LongMemEval$_\text{S}$, where dialogue histories are especially long, typed memory transforms unwieldy transcripts into traceable, citable evidence, reversing the typical failure mode of full-context inference. These reductions in input and generated tokens also translate directly to wall-clock gains, including significant drops in both median and tail latency during the answering stage.

\subsection{Practical application}
Beyond long-context stress tests, we evaluate a deployment-style scenario in healthcare, using the \emph{HealthBench} benchmark \citep{Arora2025HealthBench}, to assess whether inference-time efficiency transfers to a high-stakes domain with \emph{shorter}, multi-turn inputs. Even though the room to compress \emph{reasoning} is smaller, ENGRAM-R consistently trims generative overhead while keeping accuracy near parity, with gains on temporally anchored questions. Full results and a separate discussion appear in \textbf{Appendix~\ref{app:healthbench}}.

\section{Discussion and Conclusions}

ENGRAM-R reframes external memory not merely as a way to increase recall, but as a \textbf{mechanism for spending less compute to reach the same (or higher) fidelity}. In our setting, the practical consequence is visible in both token budgets and wall-clock time: the model reads far less, writes far less, and still answers competitively on categories/questions that require long-horizon composition with reasoning capabilities.

\paragraph{Benefits of typed representation.}
Two design decisions appear central. First, \emph{typing} (episodic, semantic, procedural) reduces competition among heterogeneous content and produces evidence sets that are semantically coherent for the downstream reasoner; this is consistent with the gains on multi-hop and temporal categories, where composition depends on keeping timelines and stable facts disentangled. Second, \emph{compact, anchored rendering} (Fact Cards) creates a low-friction interface between retrieval and inference: the model does not need to re-narrate evidence it already “has,” and the citations are reviewable by construction.

\paragraph{Limitations.}
There are several. (i) \textit{Coverage vs.\ compactness.} An aggressive evidence budget can miss rare but decisive facts; conservative budgets blunt efficiency. Budget selection remains task-dependent; thus it may be useful to have a dynamic budget. (ii) \textit{Staleness and drift.} External memory can become outdated; without refresh policies or validity intervals, models may confidently cite obsolete cards. (iii) \textit{Evaluation bias.} LLM-as-judge captures semantic agreement but is not perfect; a potential mitigation may include blinded human evaluation on LLM-graded samples.

\paragraph{Future work.}
The results suggest several immediate directions for the efficient reasoning community: (1) \textit{Adaptive budgets and routing.} Learn a small controller that sets per-query (or per-type) budgets under a global compute cap, using features from retrieval confidence, novelty, and recency. (2) \textit{Temporal consistency checking.} Equip the memory layer with lightweight temporal constraint solvers that sanity-check card sets before answering (preventing contradictions that would otherwise require long generative reconciliation). (3) \textit{Cross-modal and multilingual memory.} Extend cards to admit visual or tabular anchors and evaluate whether typed reuse equally reduces compute for multimodal LRMs.

\paragraph{Conclusions.}
Reducing test-time compute lowers the barrier to deploying reasoning-capable systems in resource-constrained environments such as edge devices, NGOs, and educational settings. In long-horizon tasks, ENGRAM-R exemplifies how structuring prior interactions into compact, typed, and citable evidence enables LRMs to \textbf{answer with fewer tokens and reduced latency}, all while maintaining high fidelity on tasks requiring complex composition. This approach offers a unique perspective on efficient reasoning, emphasizing \textbf{reuse over recompute} and demonstrating that significant computational savings can be achieved without sacrificing performance. By offering a model-agnostic, auditable mechanism that lives outside model weights, we hope to encourage broader adoption of advanced LRMs and systems involving LRMs in scenarios where computational resources are limited.

\clearpage
\bibliographystyle{plainnat}
\bibliography{engram}
\clearpage

\appendix
\section{ENGRAM Base Architecture}
\label{app:engram-base}

This appendix summarizes the \emph{base} ENGRAM system used throughout the paper (i.e., \emph{without} the fact card and citation controls introduced in Section~\ref{sec:method}). The goal is to present a compact, reproducible description of the original pipeline and data structures that were used for evaluating memory in non-reasoning models. The numbers in the subheadings correspond to labeled steps in Figure \ref{fig:engram-overview}.

\subsection{System overview}
ENGRAM converts a multi-turn dialogue into a durable, typed memory and, at query time, retrieves a small set of relevant records to condition an answering model. The pipeline comprises five components:
\begin{enumerate}[label=(\arabic*), leftmargin=1.5em]
    \item \textbf{Routing.} Decide which memory types an incoming turn should update.
    \item \textbf{Extraction \& embedding.} Normalize selected content into a record schema and compute an embedding.
    \item \textbf{Typed retrieval.} At query time, retrieve top-$k$ candidates \emph{within each} memory type by dense similarity.
    \item \textbf{Aggregation.} Merge per-type candidates, deduplicate, and truncate to a fixed budget $K$.
    \item \textbf{Prompt construction.} Serialize the selected records and inject them as context to the answering LLM.
\end{enumerate}

\subsection{Dialogue and memory notation}
Let a dialogue be $\mathcal{C}=\{x_t\}_{t=1}^{T}$, with $x_t=(s_t,u_t,\tau_t)$ denoting speaker identity $s_t\!\in\!\{A,B\}$, utterance text $u_t$, and timestamp $\tau_t\!\in\!\mathbb{R}_+$. ENGRAM maintains a typed memory state
\[
\mathcal{M}=\big(\mathcal{M}_{\mathrm{epi}},\,\mathcal{M}_{\mathrm{sem}},\,\mathcal{M}_{\mathrm{pro}}\big),
\]
to support answering queries $q$ posed after the dialogue unfolds.

\subsection{Routing and storage (1--2)}
A lightweight router maps each utterance to a compact three-bit decision
\[
r(u_t)\in\{0,1\}^3 \;\Rightarrow\; b_t=\big(b_t^{\mathrm{epi}},\,b_t^{\mathrm{sem}},\,b_t^{\mathrm{pro}}\big),
\]
indicating whether to update the \emph{episodic}, \emph{semantic}, and/or \emph{procedural} stores. For each type flagged by $b_t$, ENGRAM constructs a normalized record and pairs it with an embedding $e\in\mathbb{R}^d$ from an encoder $g:\mathcal{U}\to\mathbb{R}^d$. Records and vectors are persisted in a simple relational store (SQLite), keyed by conversation and type. The router’s 3-bit output keeps decisions interpretable and facilitates ablations.

\subsection{Typed record schemas (2)}
ENGRAM uses type-specific, minimally sufficient schemas:
\[
\begin{aligned}
m^{\mathrm{epi}} &= (t,\;\sigma,\;\delta,\;e)
\quad &&\text{(event title, brief summary, time anchor, embedding)} \\
m^{\mathrm{sem}} &= (f,\;\delta,\;e)
\quad &&\text{(fact string, time anchor, embedding)} \\
m^{\mathrm{pro}} &= (t,\;c,\;\delta,\;e)
\quad &&\text{(procedure title, normalized content, time anchor, embedding)}
\end{aligned}
\]
Typed separation reduces competition during retrieval (events vs.\ stable facts vs.\ procedures) and exposes structure that is easy to audit.

\subsection{Dense retrieval and budgeted aggregation (3--4)}
Given a query $q$, ENGRAM embeds it as $e_q=g(q)$ and retrieves candidates \emph{within each} store by cosine similarity~\citep{Karpukhin2020}
\[
R_k(q)=\operatorname{TopK}\{\mathrm{score}(e_q,m)\mid m\in\mathcal{M}_k\},\quad k\in\{\mathrm{epi},\mathrm{sem},\mathrm{pro}\}.
\]
Per-type sets are then merged and deduplicated across stores, and the union is truncated to a fixed evidence budget $K$
\[
\tilde{R}(q)=\operatorname{Truncate}_K\!\Big(\operatorname{Dedup}\big(\!\bigcup\nolimits_k R_k(q)\big)\Big).
\]
In all experiments we default to $K{=}25$, a knee point identified by a $K$-sweep ablation performed in the ENGRAM paper \citep{patel2025engram}.

\subsection{Prompt construction and answering (5)}
For multi-speaker dialogues, ENGRAM materializes speaker-specific banks $\tilde{R}(q,A)$ and $\tilde{R}(q,B)$. Each record $m$ is serialized with its temporal anchor
\[
\ell(m)=\delta(m)\;:\;\mathrm{text}(m).
\]
A deterministic formatting function $\mathrm{Template}(\cdot)$ combines the query and serialized records into the final prompt
\[
P(q)=\mathrm{Template}\!\left(q,\{\ell(m)\}_{m\in\tilde{R}(q,A)},\{\ell(m)\}_{m\in\tilde{R}(q,B)}\right),
\]
which is then passed to the answering model to obtain $\hat{a}=\mathrm{LLM}\big(P(q)\big)$. Separating banks by speaker helps preserve attribution, which is especially important during answering time.

\subsection{Remarks and scope}
This appendix describes \emph{base ENGRAM}: routing, typed storage, dense per-type retrieval, budgeted aggregation, and prompt injection of \emph{serialized snippets}. Section~\ref{sec:method} of the main paper extends the \emph{representation and control} of the retrieved set (re-rendering into compact fact cards and enforcing citation), while leaving routing, storage, and retrieval unchanged.

\subsection{Reproducibility}
To ensure transparency and reproducibility of our results, we provide the complete codebase used to implement the ENGRAM-R system, as well as all scripts required to reproduce the experiments on the LoCoMo and LongMemEval benchmarks.

The code and documentation are publicly available at the following anonymous repository:
\url{https://anonymous.4open.science/r/engram-r-7F5A/README.md}
\clearpage

\section{ENGRAM-R Architecture Ablation}
\label{app:ablation}

\paragraph{Experiment Setup.}
This ablation isolates the contribution of the fact cards and citation enforcement layer introduced in this paper. Thus we compare \textbf{ENGRAM Base} against \textbf{ENGRAM-R} and hold all components but the architecture fixed: same backbone LRM, same router, same typed stores, same dense retriever, and the same evidence budget ($K=25$). Experiments are run on LoCoMo with the QA categories used in the main results.

\begin{table}[h!]
\centering
\small
\setlength{\tabcolsep}{6pt}
\caption{LoCoMo Results: ENGRAM Base}
\label{tab:ablation_base}
\vspace{4pt}
\begin{tabular}{lrrr}
\toprule
\textbf{Category} & \textbf{Input tokens} & \textbf{Reasoning tokens} & \textbf{Judge (\%)} \\
\midrule
single-hop         & 1{,}743{,}554 & 380{,}007 & 78.3 \\
multi-hop          &   597{,}306  & 170{,}232 & 66.6 \\
temporal reasoning &   571{,}784  & 130{,}740 & 61.5 \\
open-domain        &   199{,}853  &  42{,}998 & 71.5 \\
\midrule
overall            & 3{,}112{,}497 & 723{,}997 & 73.5 \\
\bottomrule
\end{tabular}
\end{table}

\begin{table}[h!]
\centering
\small
\setlength{\tabcolsep}{6pt}
\caption{LoCoMo Results: ENGRAM-R}
\label{tab:ablation_r}
\vspace{4pt}
\begin{tabular}{lrrr}
\toprule
\textbf{Category} & \textbf{Input tokens} & \textbf{Reasoning tokens} & \textbf{Judge (\%)} \\
\midrule
single-hop         & 1{,}802{,}531 & 199{,}718 & 79.1 \\
multi-hop          &   602{,}634  &  81{,}035 & 74.5 \\
temporal reasoning &   686{,}947  &  74{,}337 & 69.2 \\
open-domain        &   201{,}366  &  23{,}334 & 54.2 \\
\midrule
overall            & 3{,}293{,}478 & 378{,}424 & 74.6 \\
\bottomrule
\end{tabular}
\end{table}

\paragraph{Findings.}
Tables~\ref{tab:ablation_base}, \ref{tab:ablation_r} show a clear pattern. Moving from ENGRAM Base to ENGRAM-R \textbf{nearly halves reasoning tokens} overall ($-47.8\%$), with small changes in input size and a modest improvement in judge accuracy (+1.1\%). The largest quality gains appear exactly where long-horizon composition is required: \textbf{multi-hop (+7.9\%)} and \textbf{temporal reasoning (+7.7\%)}, alongside sharp reductions in reasoning effort. For single-hop questions, accuracy is similar while reasoning is substantially shorter. The open-domain slice is the notable exception: while reasoning tokens shrink by $\approx$46\%, accuracy drops (–17.3\%), consistent with the need for broad background that is not captured in typed stores or that benefits from wider context.

\paragraph{Interpretation.}
ENGRAM-R trades a \emph{small, predictable} increase in input tokens (card headers/anchors and citation scaffolding) for a \textbf{large, robust decrease in generated reasoning}. The explicit instruction to \emph{cite} cards by anchor eliminates re-narration of retrieved content, and the atomic rendering reduces the need for the model to reconstruct intermediate steps already present in memory. In categories that rely on stitching together dispersed events or facts over time, this \emph{representation + control} acts like a budgeted controller: the model can answer succinctly by pointing to the right cards.

\paragraph{Takeaways.}
The ablation justifies the new components introduced in this paper:
(i) \emph{Fact Cards} convert retrieved snippets into compact, auditable units that are easy for the model to reuse; and
(ii) \emph{Citation Enforcement} turns those units into a control surface that shortens chains without degrading fidelity on long-horizon reasoning.
Under identical retrieval budgets and decoding, the combination achieves the intended goal of \textbf{reducing test-time compute at inference} while maintaining (and often improving) answer quality.
\clearpage

\section{HealthBench Evaluation}
\label{app:healthbench}

\paragraph{Motivation and setup.}
\emph{HealthBench} targets challenging, clinically relevant question answering and reasoning. We include it to test whether \textit{inference-time} efficiency mechanisms carry over to an impactful domain with domain-specific structure. We compare \textbf{full-context} against \textbf{ENGRAM-R} under the same backbone LRM, identical decoding, and the same retrieval budget $K$ as in the main experiments. Evaluation uses the same LLM-as-Judge protocol and token accounting (input vs.\ reasoning) as in Section~\ref{results}.

We evaluate on a \textbf{subset of HealthBench}, restricted to categories with \emph{multi-turn inputs}, where memory and control mechanisms are most relevant. This subset includes \textit{context-seeking}, \textit{complex-responses}, and \textit{health-data}.

\paragraph{Why HealthBench differs from LoCoMo/LongMemEval.}
Unlike \emph{LoCoMo} and \emph{LongMemEval}, HealthBench is \emph{not} designed as a long-context memory stress test. There are multi-turn inputs, but histories are \emph{much shorter} than the 10k--100k token conversations seen in our main benchmarks. Two consequences follow: (i) the fixed overhead of rendering \emph{fact cards} (anchors/headers) can slightly \emph{increase} input tokens when the original context is already compact, and (ii) the achievable reduction in \emph{reasoning} tokens is less pronounced, because there is less redundant narration to remove. We therefore expect smaller efficiency deltas than on LoCoMo/LongMemEval, with accuracy near parity if typed evidence remains sufficient.

\begin{table*}[h!]
\centering
\small
\setlength{\tabcolsep}{6pt}
\caption{\textbf{HealthBench results:} input tokens, reasoning tokens, and judge accuracy. ENGRAM-R rows are shaded light gray.\\}
\label{tab:healthbench}
\begin{tabular}{l l r r r}
\toprule
\textbf{Category} & \textbf{Setting} & \textbf{Input tokens} & \textbf{Reasoning tokens} & \textbf{LLM judge (\%)} \\
\midrule
\multirow{2}{*}{context-seeking}
  & Full-Context & 111{,}429 & 413{,}653 & 37.4 \\
  & \cellcolor{gray!9}ENGRAM-R
    & \cellcolor{gray!9}121{,}085
    & \cellcolor{gray!9}271{,}356
    & \cellcolor{gray!9}35.4 \\
\midrule
\multirow{2}{*}{complex-responses}
  & Full-Context & 120{,}964 & 310{,}730 & 32.9 \\
  & \cellcolor{gray!9}ENGRAM-R
    & \cellcolor{gray!9}133{,}423
    & \cellcolor{gray!9}246{,}423
    & \cellcolor{gray!9}32.4 \\
\midrule
\multirow{2}{*}{health-data}
  & Full-Context & 95{,}817 & 327{,}982 & 37.2 \\
  & \cellcolor{gray!9}ENGRAM-R
    & \cellcolor{gray!9}100{,}392
    & \cellcolor{gray!9}282{,}136
    & \cellcolor{gray!9}39.5 \\
\midrule
\multirow{2}{*}{overall}
  & Full-Context & 328{,}210 & 1{,}052{,}365 & 35.8 \\
  & \cellcolor{gray!9}ENGRAM-R
    & \cellcolor{gray!9}354{,}900
    & \cellcolor{gray!9}799{,}915
    & \cellcolor{gray!9}35.7 \\
\bottomrule
\end{tabular}
\end{table*}

\paragraph{Results and interpretation.}
ENGRAM-R \textbf{reduces reasoning tokens overall by roughly $\approx$24\%}. \emph{Input tokens} slightly increase (+8\% overall), and accuracy remains at parity overall (35.8\% $\rightarrow$ 35.7\%), with a notable improvement on \textbf{health-data} (+2.3\%), which benefits from explicit temporal anchors in typed memory. The mild drops on \textbf{context-seeking} and \textbf{complex-responses} suggest that, in compact settings, broader surface context can occasionally help; a hybrid policy that permits a small untyped backfill would likely recover this gap without changing the compute profile materially.

\paragraph{Takeaways for efficient reasoning.}
Even when long-context savings are structurally limited, \emph{representation and control} still reduce generative overhead: citation discourages re-narration, shrinking reasoning while keeping quality steady. In healthcare-adjacent tasks, this matters for \emph{latency} and \emph{throughput} under budget constraints.
\clearpage

\section{Full Reasoning--QA Walkthrough}
\label{app:walkthrough}

\subsection*{Worked Example.}
To illustrate the ENGRAM-R pipeline, we present a compact reasoning walkthrough.

\paragraph{Dialogue.}
The conversation unfolds as follows:
\begin{center}
\begin{tabular}{p{1.4cm}p{1.2cm}p{10.1cm}}
\toprule
\textbf{Turn} & \textbf{Speaker} & \textbf{Utterance} \\
\midrule
1 & A & ``After months of searching for a new role and packing up my old apartment, I finally \emph{moved to Seattle} last year. It took a while to adjust, but I’m starting to feel at home in the city.'' \\
2 & B & ``That’s exciting. Just don’t forget to \emph{file your tax return before April 15}---the deadline is strict and missing it could cause penalties.'' \\
3 & A & ``Appreciate the reminder. I’ve been decorating my new place, and I realized my \emph{favorite color is green}; it shows up in most of the furniture and curtains.'' \\
\bottomrule
\end{tabular}
\end{center}

At query time, the model receives the question:
\[
q = \text{``Where does A live?''}
\]

\paragraph{Pipeline Steps.}
\begin{enumerate}[leftmargin=1.8em]
    \item \textbf{Retrieval.} Relevant records are selected from episodic (relocation event), procedural (tax deadline), and semantic (favorite color) stores.
    \item \textbf{Aggregation.} The retrieved records are merged into $\tilde{R}(q)$.
    \item \textbf{Fact Card Rendering.} Each record $m \in \tilde{R}(q)$ is re-rendered into a compact representation:
    \[
    \begin{aligned}
    \phi(m_1) &= [E1,\; \text{``A moved to Seattle''},\; \tau{=}2024] \\
    \phi(m_2) &= [E2,\; \text{``B reminded A to submit tax form by Apr 15''},\; \tau{=}2024] \\
    \phi(m_3) &= [E3,\; \text{``A's favorite color is green''},\; \tau{=}2024]
    \end{aligned}
    \]
    \item \textbf{Prompt Construction.} The set $\mathcal{F}(q) = \{\phi(m)\}$ is inserted into a template with citation instructions.
    \item \textbf{Answering.} The model generates an evidence-cited answer.
\end{enumerate}

\paragraph{Reasoning Trace.}
\begin{quote}
\emph{Need to answer Q1. E1 shows A relocated to Seattle. Answer: A lives in Seattle. Cite [E1].}
\end{quote}

\paragraph{Model Output.}
\[
\hat{a} = \text{``A lives in Seattle [E1].''}
\]

\paragraph{Takeaway.}
This walkthrough shows how verbose dialogue turns are condensed into \emph{atomic, citable Fact Cards}. Instead of re-narrating evidence, the model directly cites compact evidence, yielding short and accountable reasoning traces.

\clearpage

\begin{promptbox}[title=Reasoning Prompt]
\begingroup
\small 

\setlist[itemize]{leftmargin=*, itemsep=1pt, topsep=3pt}
\setlist[enumerate]{leftmargin=*, itemsep=2pt, topsep=3pt}

\newcommand{\promptsection}[1]{\vspace{2pt}#1\par\vspace{4pt}}

You are an intelligent memory assistant with access to conversation memories and citation facts.

\promptsection{MEMORY CONTEXT (rich information)}

\promptsection{CITATION FACTS (for referencing)}

\promptsection{INSTRUCTIONS}
\begin{enumerate}[label=\arabic*., leftmargin=*, itemsep=2pt, topsep=2pt]
  \item Use the rich memory context above to understand the full situation.
  \item Answer concisely in 1--3 sentences based on the memory context.
  \item Cite supporting citation facts using \texttt{[E1]}, \texttt{[E2]} format.
  \item In your reasoning chain, use minimal tokens by:
    \begin{itemize}
      \item ALWAYS reference the question as “Q1” (never repeat the full question).
      \item ALWAYS cite facts by label only (e.g., “E1 shows…”, “E2 indicates…”).
      \item NEVER repeat full fact content in reasoning.
      \item Be extremely concise and focused.
    \end{itemize}
  \item Only say “not enough info” if truly no relevant information exists.
\end{enumerate}

\medskip

\promptsection{REASONING EXAMPLE}
\emph{Need to answer Q1. E1 shows Melanie ran charity race on May 20. E2 indicates Caroline was proud. Answer: May 20, 2023. Cite [E1].}

\medskip
\promptsection{QUESTION TO ANSWER}
Q1: Where does A live?

\medskip
\promptsection{Please provide your answer with reasoning}
\endgroup
\end{promptbox}

\end{document}